\documentclass[aps,prl,showpacs,twocolumn,groupedaddress]{revtex4}

\usepackage{graphicx}
\usepackage{dcolumn}
\usepackage{bm}

\begin{document}


\title{Gate-Voltage Control of Chemical Potential and Weak Anti-localization in Bi$_2$Se$_3$ }
\author{J.~Chen\,$^*$, H.~J.~Qin\,\footnote{These two authors contributed equally to this work.}, F.~Yang, J.~Liu, T.~Guan, F.~M.~Qu, G.~H.~Zhang, J.~R.~Shi, X.~C.~Xie, C.~L.~Yang, K.~H.~Wu\,\footnote{Correspondence author for MBE growth, Email: khwu@iphy.ac.cn},Y.~Q.~Li\,\footnote{To whom all other correspondences should be addressed, Email: yqli@iphy.ac.cn}, and L.~Lu}
\affiliation{Institute of Physics, Chinese Academy of Sciences, Beijing 100190, China}

\date{\today}

\begin{abstract}
We report that Bi$_2$Se$_3$ thin films can be epitaxially grown on SrTiO$_{3}$ substrates, which allow for very large tunablity in carrier density with a back-gate. The observed low field magnetoconductivity due to weak anti-localization (WAL) has a very weak gate-voltage dependence unless the electron density is reduced to very low values. Such a transition in WAL is correlated with unusual changes in longitudinal and Hall resistivities. Our results suggest much suppressed bulk conductivity at large negative gate-voltages and a possible role of surface states in the WAL phenomena. This work may pave a way for realizing three-dimensional topological insulators at ambient conditions.

\end{abstract}

\maketitle

A new state of matter, coined as topological insulator \cite{Qi10,Hasan10}, was recently discovered to exist in three dimensions \cite{Fu07a,Hsieh08}. It is characterized by an insulating bulk and conducting surface states of massless helical Dirac fermions as a consequence of strong spin-orbit coupling \cite{Fu07b,Moore07,Qi08,Schnyder08}. The surface of the topological insulator can be viewed as a novel type of two-dimensional electron system (2DES) that may be a fertile ground for exploring exciting new physics as well as possible applications in spintronics and quantum information \cite{Hasan10}. Many of them are not possible in its topologically trivial predecessors such as 2DES based on semiconductor heterostructures and graphene. Among all known materials that could host the topological surface states, bismuth selenide (Bi$_2$Se$_3$) is most attractive because of a large band gap ($\sim 0.3$\,eV) in the bulk and a single Dirac cone in the surface energy spectrum \cite{ZhangH09,Xia09}.


Despite numerous theoretical proposals that attempt to exploit their exotic properties, the 3D topological insulators have so far only confirmed to exist in ultra-high vacuum (UHV) environment \cite{Hsieh08,Xia09,Chen09,Hsieh09b,ZhangT09,ZhangY09}. All of previously reported transport studies have been bothered by a conducting bulk \cite{Taskin09,Checkelsky09,Peng10,Analytis10}.
Even under circumstances of UHV, suppression of the bulk conductivity appears to be a challenging task. Hsieh et al.\ reported that the chemical potential lies in the bulk band gap for a freshly cleaved, Ca-doped Bi$_2$Se$_3$ crystal, and an 18-hour stay in UHV would lift the chemical potential into the bulk conduction band \cite{Hsieh09a}. Analytis et al.\ observed that an exposure of a cleaved Bi$_2$Se$_3$ surface in the air for 10\,seconds is sufficient to introduce a significant amount of electrons into the bulk \cite{Analytis10}.
It was demonstrated that the surface doping effect can be overcome by tuning chemical potential with NO$_2$ molecules deposited onto the Bi$_2$Se$_3$ surface in UHV so that a true topological insulator was obtained \cite{Hsieh09a}. This method is however cumbersome, if not impossible, for preparing devices suitable for transport studies. Therefore, it is very important to find more convenient means to lower the chemical potential in order to bring the surface states into the so-called topological transport regime, in which suppression of the bulk conductivity is a prerequisite.

Here we report that large tunability in the chemical potential in Bi$_2$Se$_3$ is achieved with a back-gate. The gate-voltage dependencies of longitudinal and Hall resistivities as well as the weak anti-localization part of magnetoconductivity coherently point to a much depleted bulk and possible relevance of the surface states.

The Bi$_2$Se$_3$ films were grown on SrTiO$_{3}$(111) substrates with molecular beam epitaxy. Fig.\,1a shows an \textit{in situ} scanning tunneling microscopy (STM) image of a Bi$_2$Se$_3$ thin film. It displays many triangular-shaped flat terraces with a step height close to 1\,nm, corresponding to one quintuple-layer (QL).  The QL-by-QL growth mode of Bi$_2$Se$_3$ \cite{ZhangG09} enabled us to precisely control the film thickness. A 50\,nm thick Pd film was subsequently deposited onto the back of the substrate to serve as a back-gate. The Bi$_2$Se$_3$ films were then patterned into Hall bars of millimeter sizes. The electron transport measurements were carried out in a cryostat with $B$ up to $7$\,T and at temperature $T=1.8$\,K, unless otherwise specified. Hall measurements showed that undoped Bi$_2$Se$_3$ films are of n-type. We also found that doping the films with Ca can lower the electron density, consistent with previous work on doped bulk materials \cite{Hor09}. The data to be presented here were mainly taken from two samples. Sample A is a 48\,nm thick undoped Bi$_2$Se$_3$ film, whereas sample B is doped with Ca and $10$\,nm thick.

\begin{figure}
\centering
\includegraphics*[width=6.0 cm]{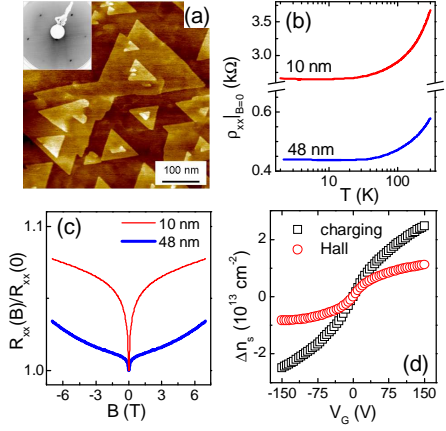}
\caption{(color online) (a) STM image and low energy electron diffraction pattern (inset) of a Bi$_2$Se$_3$ single crystalline thin film grown on a SrTiO$_{3}$(111) substrate; (b) $T$-dependence of $\rho_\mathrm{xx}$ of samples A (lower) and B (upper) at $B=0$; (c) Normalized magnetoresistance of samples A (lower) and B (upper); (d) Relative changes in carrier density with respect to $V_G=0$ extracted from Hall measurements (circles) and charging effects (squares).} \label{Fig1}
\end{figure}

As shown in Fig.\,1b, the longitudinal resistivities at $B=0$ ($\rho_\mathrm{xx}^0$) of samples A and B have similar temperature dependence.
When a perpendicular magnetic field is applied, both samples show positive magnetoresistance (MR) with a cusp-like minimum at $B=0$.  The MR of sample A in large fields has a parabolic $B$-dependence, which is not observed in sample B. The Hall resistivities ($\rho_\mathrm{xy}$) of both samples varies linearly with $B$ without any noticeable curvature. The electron densities extracted from the Hall coefficients are $n_s^\mathrm{Hall}=-1/(e R_H)\approx$ 3.7 and 1.1 $\times 10^{13}$\,cm$^{-2}$ for samples A and B, respectively.

The electron density can be modified substantially by applying the gate voltage $V_G$. Fig.\,1d shows the relative changes in the carrier density in sample B with respect to $n_s$ at $V_G=0$. They were measured with a charging experiment, in which a constant current of 1\,nA was used to charge the Bi$_2$Se$_3$/STO/Pd capacitor. The recorded charging time during the change of $V_G$ thus provided a direct measurement of the variation in the \textit{total} carrier density in the Bi$_2$Se$_3$ electrode. It has a nonlinear dependence on $V_G$, which is a typical dielectric response of STO single crystals to the applied electric fields \cite{Christen94}.  The relative change in the total carrier density with respect to $n_s$ at $V_G=0$, $\Delta n_s$, reaches $\pm2.5\times10^{13}$\,cm$^{-2}$ at $V_G=\pm150$\,V, demonstrating a superb capability in tuning the carrier density.

As shown in Fig.\,1d, the relative changes in $n_s$ extracted from the Hall measurements, $\Delta n_s^\mathrm{Hall}$, are significantly smaller in magnitude than $\Delta n_s$. This suggests that the Hall measurements give underestimated $n_s$ values, even though $\rho_\mathrm{xy}$ has a perfectly linear dependence on $B$. This is not surprising because multiple types of charge carriers are very likely to coexist in the Bi$_2$Se$_3$ film. According to the photoemission studies \cite{Xia09,Hsieh09a,Analytis10,ZhangY09}, the energy difference between the conduction band minimum and the Dirac point of the surface states is only about 0.2\,eV. This means that the two surfaces in the topological regime can only hold up to about $0.5\times10^{13}$ electrons/cm$^2$. Therefore, even in sample B, which has a very low density, there exist a large number of conduction band electrons at $V_G=0$. Considering that the bottom and top surfaces have different interfaces, we expect at least three different types of charge carries in the device, namely the bulk electrons and the carriers related to the top and the bottom surface states. Under this circumstance, the Hall coefficients at the low $B$ limit ($\omega_c\tau\ll 1$, which is satisfied in both samples) do not provide accurate measurement of the \textit{total} carrier density. It rather gives values in which more mobile charge carries have more contribution.

\begin{figure}
\centering
\includegraphics*[width=8 cm]{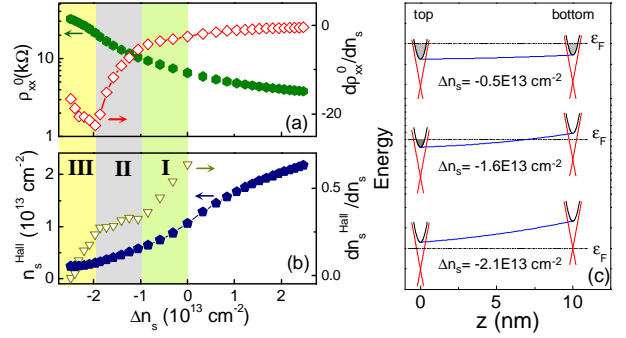}
\caption{(color online) Plotted as a function of $\Delta n_s$ are (a) $\rho_\mathrm{xx}^0$ (hexagons) and its derivative (diamonds); (b) $n_s^\mathrm{Hall}$ (pentagons) and its derivative (triangles). The negative gate voltages are partitioned into three zones (I, II, III) according to the degree of carrier depletion. (c) Calculated band diagrams based on a simple model described in the text.} \label{Fig2}
\end{figure}

Fig.\,2 shows the carrier density dependence of longitudinal resistivity $\rho_\mathrm{xx}^0$ and $n_s^\mathrm{Hall}$. When $n_s$ is lowered, $\rho_\mathrm{xx}^0$ increases monotonically, but its derivative, $d\rho_\mathrm{xx}^0/dn_s$, has a minimum at $\Delta n_s\approx-1.9\times10^{13}$\,cm$^{-2}$, where the slope of $d\rho_\mathrm{xx}^0/dn_s$ changes its sign from positive to negative \textit{abruptly}. Unusual features are also present in $n_s^\mathrm{Hall}$. The derivative $dn_s^\mathrm{Hall}/dn_s$ does not vary smoothly with $n_s$. It rather shows sharp transitions at about $-1.0$ and $-1.9\times10^{13}$\,cm$^{-2}$. The latter coincides with the transition in $d\rho_\mathrm{xx}^0/dn_s$.

Based on these observations, we divide the negative $\Delta n_s$ region in Fig.\,2 into three zones to represent different degrees of carrier depletion. In zone I, which spans from 0 to $-1.0\times10^{13}$\,cm$^{-2}$, $\rho_\mathrm{xx}^0$ varies very slowly. The magnitude of $d\rho_\mathrm{xx}^0/dn_s$ increases rapidly in zone II ($-1.9<\Delta n_s<-1.0$ $\times10^{13}$\,cm$^{-2}$). This is normally what one would expect from depleting a typical semiconductor layer. At sufficiently low densities, metal-insulator-transition should take place in the bulk and the rapid increase in $|d\rho_\mathrm{xx}^0/dn_s|$ in zone II may be a precursor for such a transition. When $\Delta n_s<-1.9\times10^{13}$\,cm$^{-2}$ (zone III), further depletion leads to the suppression of the bulk conductivity and hence the transport is dominated by the surface (or interface) states. This could account for the sudden change in the slope of $d\rho_\mathrm{xx}^0/dn_s$. Another noteworthy feature in this deep depletion regime (zone III) is that $n_s^\mathrm{Hall}$ only varies from $-0.31$ to $-0.24$ $\times10^{13}$\,cm$^{-2}$. Its slope becomes nearly zero when $\Delta n_s$ approaches to the lower limit. This would be difficult to explain if every part of the sample still remained n-type. A likely scenario is that the chemical potential at the bottom surface has dropped below the Dirac point.

The profile of the chemical potential for a given $\Delta n_s$ can be calculated by solving the 1D Poisson equation. For the simplest case, we do not consider band bending near the top and bottom surfaces at $V_G=0$. If we assume a bulk electron density of $1\times10^{19}$\,cm$^{-3}$, $\Delta n_s\approx-2\times10^{13}$\,cm$^{-2}$ is then required to reach the situation in which the chemical potential on the top surface is lowered to the conduction band minimum. Fig.\,2c displays three band diagrams calculated for $\Delta n_s=-0.5$, $-1.6$ and $-2.1\times10^{13}$\,cm$^{-2}$, which may provide a crude picture for the evolution of the chemical potential and the carrier depletion as $V_G$ decreases. Many experimental factors, for instance, the band bending effects near the top and bottom surfaces as well as impurity levels, can certainly modify the results. Nevertheless, the basic trend shown in Fig.\,2c should remain intact.

Fig.\,3 shows a detailed gate voltage dependence of magnetoconductivity of samples A and B. It is defined as $\Delta\sigma(B)=\sigma_\mathrm{xx}(B)-\sigma_\mathrm{xx}|_{B=0}$. For both samples $\Delta\sigma(B)$ has a cusp-like maximum at $B=0$, which is typical for the weak anti-localization. For sample A (Fig.\,3a), within the range of all accessible carrier densities, the low field part of $\Delta\sigma(B)$ varies very little. At $B>1$\,T, $\Delta\sigma(B)$ decreases as $V_G$ becomes smaller. The high field part has a parabolic shape, which probably originates from semiclassical effects.

\begin{figure}
\centering
\includegraphics*[width=7.0 cm]{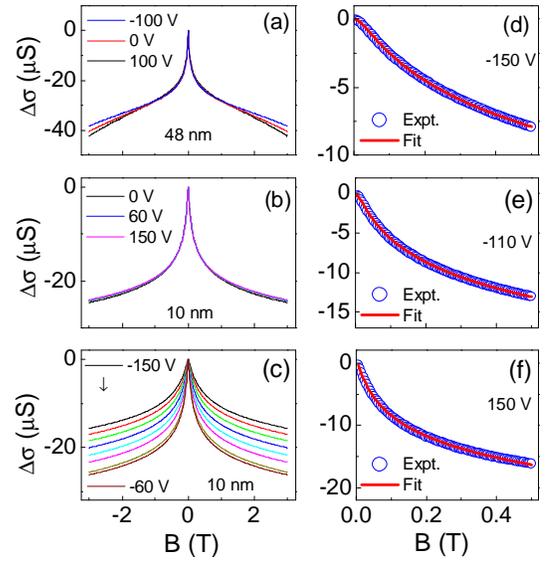}
\caption{(color online) Magnetoconductivity, defined as $\Delta\sigma(B)=\sigma_\mathrm{xx}(B)-\sigma_\mathrm{xx}|_{B=0}$ of (a) sample A; (b) sample B at $V_G=$0, 20, 40, 60, 100, and 150\,V; (c) sample B at $V_G=-$150\,V, $-140$, $-130$, $-120$, $-110$, $-100$, $-80$, and $-60$\,V (from top to bottom). Fitted curves (lines) with Eq.\,(1) are also shown in (d) $V_G=-150$\,V, (e) $-110$\,V, and (f) $150$\,V for sample B.
} \label{Fig3}
\end{figure}

The gate voltage dependence of the magnetoconductivity of sample B is shown in Figs.\,3b and 3c. For positive gate voltages, all $\Delta\sigma(B)$-$B$ curves are nearly identical, even though there are more than a factor of 2 change in $R_H$ and an increase of $2.5\times10^{13}$\,cm$^{-2}$ in $n_s$. In contrast, when $V_G$ is lowered to $-60$\,V and below, $\Delta\sigma(B)$ varies drastically with $V_G$. Measurements on several other samples with various carrier densities that spread between those of samples A and B further confirm the following observation: the back-gate voltage can induce significant amount of change in the low field magnetoconductivity only when the carrier density is extremely low.

In case of coexistence of a bulk Fermi surface with the surface states, the Bi$_2$Se$_3$ thin films may be treated as traditional 2D electron systems with strong spin-orbit interaction. Hikami-Larkin-Nagaoka (HLN) equation \cite{Hikami80} is often applied to account for the magnetoconductivity in this kind of systems. In the limit of very strong spin-orbit interaction and low mobility, i.e.\ $\tau_\phi\ll$ $\tau_\mathrm{so}$ and $\tau_\phi\ll \tau_\mathrm{e}$, the HLN equation is reduced to
\begin{equation}
\label{eq:WALtopo}
\Delta\sigma(B)\simeq
-\alpha\cdot\frac{e^2}{2\pi^2\hbar}
\left[
\psi\left(\frac{1}{2}+\frac{B_\phi}{B}\right)
-\ln\left(\frac{B_\phi}{B}\right)
\right],
\end{equation}
where $\tau_\mathrm{so}$ ($\tau_e$) is the spin-orbit (elastic) scattering time, $\alpha=1/2$, $\psi$ is the digamma function, and $B_\phi=\hbar/(4De\tau_\phi)$ is a characteristic field related to the dephasing time $\tau_\phi$. Here $D$ is the diffusion constant.
For the surface states in the topological regime, to the best of our knowledge, no theory has yet been given to quantitatively describe the weak anti-localization arising from the Berry's phase of the Dirac fermions \cite{Fu07a,Roushan09}. We notice, however, that McCann et al.\,\cite{McCann06} developed a theory for graphene, in which WAL could be suppressed by intervalley scatterings as well as chirality-breaking intravalley scatterings by lattice defects (e.g.\ ripples and dislocations). These effects are absent in the single-cone spin-helical system encountered here, so we expect that the magnetoconductivity of \textit{one} topological surface of Bi$_2$Se$_3$ follows the same form as Eq.\,(1) with $\alpha=1/2$~\footnote{The coefficient $\alpha$ would be $4\times1/2=2$ for an infinite graphene sheet free of imperfections.}. Our fitting results with various samples showed that the conditions $\tau_\phi\ll$ $\tau_\mathrm{so},$ $\tau_\mathrm{e}$ are indeed satisfied at $T=1.8$\,K. As shown in Figs.\,3d-f, Eq.\,(1) does provide reasonably good fits to the data at all gate voltages.

\begin{figure}
\centering
\includegraphics*[width=7.0 cm]{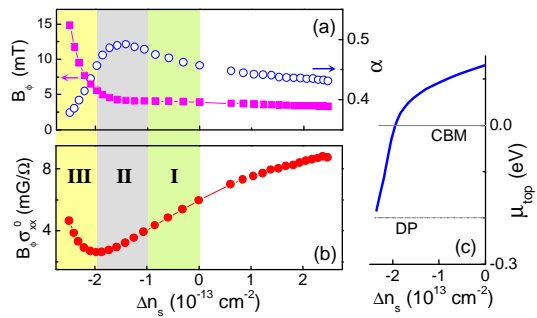}
\caption{(color online) (a) $B_\phi$ (solid squares) and $\alpha$ (open circles), (b) $B_\phi\sigma_\mathrm{xx}^0$, as a function of $\Delta n_s$. Here $\sigma_\mathrm{xx}^0$ is the conductivity at $B=0$. The partition of the negative $V_G$ region is same as that in Fig.\,2. (c) Calculated chemical potential on the top surface with respect to the conduction band minimum (CBM). The Dirac point (DP) is located at $-0.2$\,eV.} \label{Fig4}
\end{figure}

Fig.\,4a displays detailed fitting results for sample B. $B_\phi$ varies very little for a wide range of densities, i.e.\ $n_s=-1.4$ to $+2.5\times10^{13}$\,cm$^2$. When $n_s$ becomes lower (in zone II), $B_\phi$ starts to vary appreciably with $V_G$. It is striking that $B_\phi$ nearly triples in a  very narrow range of carrier densities in zone III.
The correlation between the WAL data and $\rho_\mathrm{xx}$ is shown more clearly in the $B_\phi\sigma^0_\mathrm{xx}$-$\Delta n_s$ plot (Fig.\,4b), which has a minimum at the boundary between zones II and III.

As shown in Fig.\,4a, the coefficient $\alpha$ extracted with Eq.\,(1) is close to $1/2$ for all carrier densities. One possible explanation is that the Bi$_2$Se$_3$ thin film can be treated as a 2D system described by the HLN equation, but it is difficult to reconcile with the fact that $B_\phi$ increases by less than 15\% for a wide range of densities ($\Delta n_s=-1.4$ to $2.5\times10^{13}$\,cm$^{-2}$), while $\rho_\mathrm{xx}$ and $n_s^\mathrm{Hall}$ are changed by a factor of 3.4 and 4.7, respectively. Such large changes are usually reflected in $D$ and $\tau_\phi$, and $B_\phi$ would change accordingly.

An alternative explanation might come from treating the top and bottom surface states as well as the bulk electrons separately. Since both surfaces were measured simultaneously in the experiment, this would lead to $\alpha=1$ if they contribute equally. Apparently this is not the case. Considering that $V_G$ is applied from the back of the sample, the electron states at and near the bottom surface will be influenced before any impact can be exerted on the top surface (see Fig.\,2c). The weak carrier density dependence of $B_\phi$ for such a wide range of $V_G$ implies that the bottom surface or the nearby bulk does not contribute much to the magnetoconductivity at low fields. It rather mainly originates from the electron states at or near the top surface. This may be understood as follows. The bottom surface has many defects because of the lattice mismatch with the STO substrate. Due to the layered nature of the Bi$_2$Se$_3$ crystal and lattice relaxation during the growth, the top surface is expected to be of considerably better quality. Consequently, the top surface has much higher carrier mobility ($\mu$) and thus smaller $B_\phi$ since $D\propto\mu$. This could result in negligible contribution from the bottom surface. In addition, the Bi$_2$Se$_3$ crystal has inversion symmetry, which may account for insignificant contribution from the bulk.

If the low field magnetoconductivity is indeed mainly contributed by the states at or near the top surface, the large increase $B_\phi$ in zone III can be explained qualitatively. In this low density regime, we anticipate a significant drop in the local chemical potential at the top surface as $V_G$ becomes more negative (Fig.\,4c). This will result in substantial decrease in the carrier density of the top surface states. For a 2D system the diffusion constant is $D=1/2v_F^2\tau_e$, where the Fermi velocity $v_F$ is constant for Dirac fermions. The larger $B_\phi$ at lower densities can therefore be attributed to reduced screening and stronger electron-electron interaction effects, which are known to shorten $\tau_e$ and $\tau_\phi$. In case that the conduction band electrons are not fully depleted near the top surface, similar arguments can also be applied. The corresponding $n_s$-dependence of $B_\phi$ should, however, be \textit{qualitatively} different because the dimensionality of the electron system is changed from 2D to 3D (or quasi-2D). The $B_\phi\sigma_\mathrm{xx}^0$ minimum observed near the boundary between zones II-III in Fig.\,4b might be associated with a transition of the electron states from a bulk-like to a more surface-dominating type.

Finally, we briefly remark on the coefficient $\alpha$, which has a sizable, density-dependent deviation from $1/2$. This is not expected from the theories mentioned above. Possible explanations may arise from electron-electron interaction effects \cite{Lee85} as well as states related to defects in Bi$_2$Se$_3$ \cite{Checkelsky09}. Furthermore, the large electric field induced by the back-gate voltage might significantly modify the spin-orbit interaction and hence spin relaxation via the Rashba effects \cite{ZhangY09,Knap96}. A thorough explanation of the data is out of the scope of this paper, but it certainly deserves further work along this direction. It will be of great interest to see whether such studies could provide a new method to distinguish a topological metal from an ordinary 2D electron system.

We are grateful to D.~Maryenko, K.~Chang, N.~Cooper, X.~Dai, and Z.~Fang for stimulating discussions. We acknowledge financial support from Ministry of Science \& Technology of China, National Science Foundation of China, and Chinese Academy of Sciences.


\end{document}